\documentclass[12pt]{iopart}

\usepackage{iopams}
\usepackage{graphicx}
\begin{document}

\title{Asymmetric Fermi superfluid in a harmonic trap}

\author{C.-H. Pao$^1$ and S.-K. Yip$^2$}

\address{$^1$Department of Physics, National Chung Cheng University, Chiayi 621,
Taiwan}
\address{$^2$Institute of Physics, Academia
Sinica, Nankang, Taipei 115, Taiwan} \ead{pao@phy.ccu.edu.tw} 

\begin{abstract}
We consider a dilute two-component atomic fermion gas with unequal
populations in a harmonic trap potential using the mean field theory
and the local density approximation. We show that the system is
phase separated into concentric shells with the superfluid in the
core surrounded by the normal fermion gas in both the weak-coupling
BCS side and near the Feshbach resonance. In the strong-coupling BEC
side, the composite bosons and left-over fermions can be mixed. We
calculate the cloud radii and compare axial density profiles
systemically for the BCS, near resonance and BEC regimes.

\end{abstract}

\submitto{\JPCM}
\maketitle

\section{\label{intro}Introduction:}
The original BCS state for superconductors considers pairing between
two species of fermions with equal populations. For a long time,
theorists studied the fermion system with unequal species, or
mismatched Fermi surfaces, and proposed this system may have
different ground state \cite{FFLO}, in particular the so called
Fulde-Ferrell-Larkin-Ovchinnikov (FFLO) phase.
 Experimentally, however,
such superfluid states remain unclear because of the difficulty in
preparing the magnetized superconductors.

Experiments with ultra-cold atoms have opened a new era to study
this fermion system with unequal populations. Through the Feshbach
resonance \cite{Feshbach}, the effective interaction between atoms
can be varied over a wide range such that the ground state can be
turned from a weak-coupling BCS superfluid to a strong-coupling
Bose-Einstein condensation (BEC) regime. In the homogeneous system,
theoretical studies \cite{Carlson05,pao06,Son05} of the unequal
fermion species show that the phase transition must occur when the
resonance is crossed, in contrast to the equal population case where
a smooth crossover takes place \cite{EL,SRE93}. Breached pair phase
\cite{LW}, phase separated states are also proposed
\cite{Bedaque03,Caldas04} in this system.

Two recent experiments \cite{ketterle06,hulet06} studied the trapped
$^6$Li atoms with imbalanced spin populations and obtained the
density profiles for various population differences. Both groups
found the system contains a superfluid core surrounded by a normal
fermions and provide evidence for phase separation near the
crossover.


In this paper, we study this imbalanced fermion system by mean field
approximation and evaluate the density profiles for various coupling
strengths from weak-coupling BCS superfluid to strong coupling BEC
regime. In particular, we calculate the axial density profiles,
superfluid and minority cloud radii, and distinguish between the
phase separation and Bose-Fermi mixture regimes.
 This
paper is organized as follows: In Sec. II we briefly review the
mean-field approximation for the dilute fermion atoms with unequal
populations.
In Sec. III 
we present our results for various polarizations from
weaking-coupling BCS superfluid to strong-coupling BEC side. We show
that the axial density profiles are constant within the superfluid
core and decrease beyond the phase boundary for the phase
separations but are smoothly decreasing functions for entire trap
for the mixtures. Finally, we conclude with a briefly summary in
Sec. IV.

While this work was in progress, several theoretical papers have
also studied the same problem under similar approximations
\cite{Yi,deSilva,Haque,chevy06} or going beyond
 \cite{kinnunen05,pieri05}.
Basically, ref \cite{Yi,deSilva,Haque,chevy06,kinnunen05} also
conclude the system is phase separated into concentric shells with
superfluid in the center and surrounding by leftover fermions near
the resonance. The strong-coupling BEC limit has also been studied
\cite{Yi,deSilva,pieri05}. In this case, the composite bosons and
unpaired fermions can mix. As the population difference increases
the unpaired fermions can even penetrate into the superfluid core.
Our paper provides a more systematic study of the entire BCS-BEC
regimes for all polarizations.

\section{\label{form}Formalism:}
Restricting ourselves to wide Feshbach resonance,
 the two-component fermion system
can be described by an effective one-channel Hamiltonian
\begin{equation}
H\ =\ \label{eqh} \sum_{{\bf k}, \sigma} \xi_\sigma ({\bf k})
c^\dagger_{{\bf k},\sigma} c_{{\bf k},\sigma}\ +\ g \sum_{\bf
k,k^\prime, q} c^\dagger_{{\bf k+q},\uparrow} c^\dagger_{{\bf
k^\prime -q},\downarrow} c_{{\bf k^\prime},\downarrow} c_{{\bf
k},\uparrow}\ ,
\end{equation}
where $\xi_{\sigma}({\bf k})\, =\, \hbar^2 k^2/2m - \mu_\sigma$ and
the index $\sigma$ runs over the two spin components. Within the BCS
mean field approximation at zero temperature, the excitation
spectrum in a homogeneous system for each spin is (see e.g.
\cite{WY03} for details)
\begin{equation}
E_\sigma ({\bf k})\, = \, \frac{ \xi_{\sigma}({\bf k}) - \xi_{-
\sigma}({\bf k})}{ 2}\ +\ \sqrt{ \left ( \frac{ \xi_{\sigma}({\bf
k}) + \xi_{-\sigma} ({\bf k})} {2 } \right )^2\, + \, \Delta^2}\ ,
\label{eqdisp}
\end{equation}
where $\xi_{\sigma}({\bf k})\, =\, \hbar^2 k^2/2m - \mu_\sigma$ are
the quasi-particle excitation energies for normal fermions, and
$-\uparrow \equiv \downarrow$. For an inhomogeneous system, $e.g.$
the system in a harmonic trap, a finite system sized effect should
be included \cite{bruun99}. However, the system can be treated as
homogeneous locally if the number of particles are sufficiently
large.
 A local density approximation, or
Thomas-Fermi approximation (TFA), is applied and the chemical
potential for spin $\sigma$ is replaced by
\begin{equation}
\mu_\sigma (r) \ =\ \mu^0_\sigma\ - \frac{1}{2} m \omega^2 r^2\ ,
\end{equation}
with $\omega$ the isotropic trap frequency and $r$ the distance from
the trap center. We shall show results explicitly only for the
isotropic trap. In the local density approximation, the densities
$n_\sigma ({\bf r})$ depends only on the local chemical potentials
$\mu_\sigma({\bf r})$. Hence the density profile of an anisotropic
trap can be related to an isotropic one by rescaling the spatial
coordinates appropriately. We then introduce the average chemical
potential
\begin{equation}
\mu (r)\ \equiv\ \frac{1}{2} [\mu_\uparrow (r) + \mu_\downarrow
(r)]\ =\ \mu_0\ -\ \frac{1}{2} m \omega^2 r^2\ ,
\end{equation}
and the difference $h\ \equiv \ [\mu_\uparrow (r) - \mu_\downarrow
(r) ]/2\, =\, (\mu^0_\uparrow - \mu^0_\downarrow)/2 $. The
dispersion relation in Eq. (\ref{eqdisp}) becomes
\begin{equation}
E_{\uparrow,\downarrow} ({\bf k,r})\, = \, \sqrt{ \xi({\bf k,r})^2 +
\Delta^2({\bf r})} \mp h \label{Ek}
\end{equation}
where $\xi({\bf k,r}) \equiv \hbar^2 k^2/2m - \mu (r)$. We take spin
up to be the majority species so that $h$ and $E_\uparrow$ are
positive always. Then the density profiles in a harmonic trap are
\begin{eqnarray}
n_s(r)& = & n_\uparrow(r) + n_\downarrow (r)\ =\ \int { d^3 k \over
(2 \pi)^3} \left [ 1 - {2 \xi({\bf k, r}) \over E_\uparrow +
E_\downarrow} f(-E_\uparrow)\right  ]\ , \label{eqns}\\
n_d (r) & = & n_\uparrow(r) - n_\downarrow (r)\ =\ \int { d^3 k
\over (2 \pi)^3} f(E_\uparrow)\ , \label{eqnd}
\end{eqnarray}
and the total number of particles $N\, = \, \int d^3 r n_s(r) $.
Here $f$ is the Fermi function. The polarization of the system is
defined as
\begin{equation}
P \ \equiv\ {N_\uparrow - N_\downarrow \over N}\ =\ {1 \over N}\int
d^3 r n_d(r)\ .
\end{equation}

Now the pairing field $\Delta$ depends on position also. In the
local density approximation, it obeys an equation similar to the
homogeneous case \cite{pao06,WY03}:
\begin{equation} - \frac{ m }{ 4 \pi a} \Delta (r)\, =\, \Delta(r)
 \int \frac{ d^3 k}{ (2\pi)^3}
\left [ \frac{ 1 - f(E_\uparrow) - f(E_\downarrow) }{ E_\uparrow +
E_\downarrow }\, -\, \frac{ m }{ \hbar^2 k^2} \right ]\ .
\label{eqgap}
\end{equation}
For a given scattering length $a$, we solve equations (\ref{eqns}),
(\ref{eqnd}) and (\ref{eqgap}) self-consistently for fixed total
number of particles $N$ and polarization $P$. The solutions to
the``gap equation'' (\ref{eqgap}) may not be unique. The physical
solution is determined by the condition of minimum free energy among
the multiple solutions. We describe the detail procedures in
\ref{freeenergy}.

\section{\label{result}Results and Discussions}

In this section, we investigate the density profiles for various
polarizations and coupling strengths from positive detuning BCS
superfluid to negative detuning BEC side. With the aid of density
profiles, we evaluate the radii of the superfluid phase boundaries
for various cases and compare to the current experimental results.
We close this section with a discussion of axial density profiles.
Phase separation versus Bose-Fermi mixture can be clarified through
the axial density profiles of the population difference.

\begin{figure}
\includegraphics{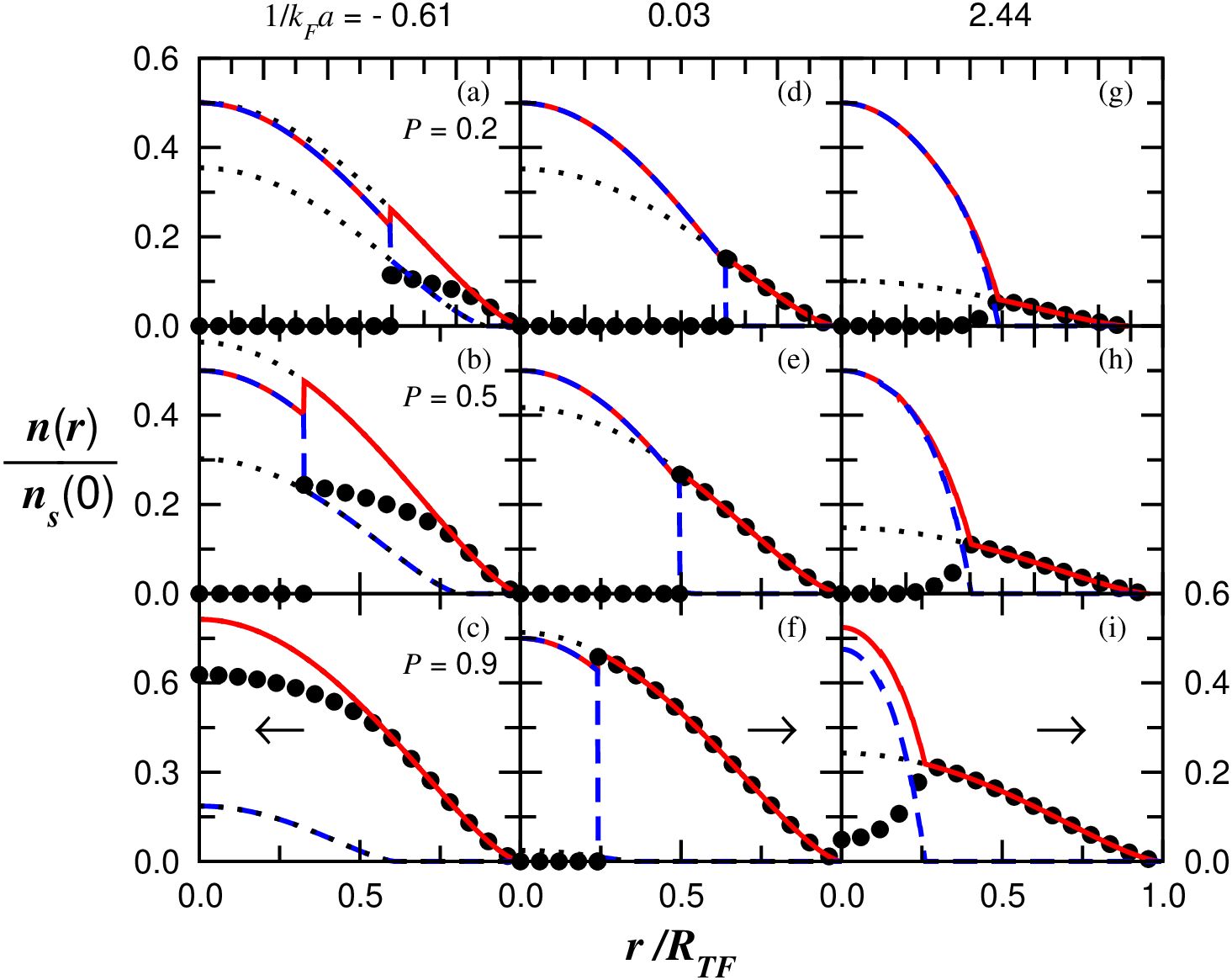} \vspace*{5.8in} \caption{(color online). Radial
density profiles for $n_\uparrow(r)$ (solid lines),
$n_\downarrow(r)$ (dashed lines), and $n_d (r)$ (solid circles). The
dotted lines are the normal state density profile with the same
density as in the normal region. The coupling strengths (each
column) $1 / (k_F a) = $ -0.61 (left, BCS-side), 0.03 (middle, near
resonance), 2.44 (right, BEC-side) and the polarizations (each row)
$P = $ 0.2, 0.5, and 0.9. The total number of fermions are fixed to
$N = 2.0 \times 10^5$. $R_{TF}$ is the size of the Thomas-Fermi
radius of the normal spin-up cloud.} \label{fig1}
\end{figure}

In Fig. \ref{fig1}, we plot the radial density profiles for three
different coupling strengths $1 / k_F a\, =\, -0.61, \, 0.03$, and
$2.44$ (for different columns) with polarization $P\, =\, 0.2\, ,
0.5$, and $0.9$ (for different rows). The total number of particles
are fixed to $2 \times 10^{5}$. Here $k_F$ is the Fermi wavevector
at the trap center for an ideal symmetric Fermi gas with the same
total number of particles. In all these plots, the system shows a
superfluid cloud surrounded by a normal Fermi gas except the case in
Fig. \ref{fig1}(c) where the system is completely in the normal
state with the polarization $P = 0.9$. It is consistent with the
experimental observation \cite{ketterle06} that the superfluid is
destroyed by a sufficiently large population difference. We remark
here however that the critical population difference $P_c$ for the
destruction of superfluidity obtained here is much larger than that
found in the experiment \cite{ketterle06}. For $1/k_F a = -0.61$
here, $P_c > 0.6$ theoretically whereas an extrapolation of the data
of \cite{ketterle06} gives $P_c < 0.3$ (see further discussions
below). For less polarized system on the BCS side [Fig.
\ref{fig1}(a) and (b)], it shows a clear phase separation between
the superfluid and a normal Fermi gas. Note that within the
superfluid cloud, the population difference is zero and the system
is just like the typically unpolarized superfluid. Outside the
superfluid cloud, both components of the fermions exist in the
normal state which indicates a fraction of the fermions which are
not paired-up even at the zero temperature. The density profile of
the population difference $n_d(r)$ peaks at the superfluid phase
boundary and decreases gradually toward the edge of the trap. Its
value is equal to the density profile of the majority when $r \ge
r_\downarrow$,
the radii of the minority cloud. At the superfluid phase boundary,
both the majority and minority density profiles exhibit
discontinuous which has been observed also by the others
\cite{Yi,deSilva,Haque}.

Near resonance [Fig. \ref{fig1}(d)-(f)], similar phase separated
states are observed. However, most of the minority are paired up in
this regime such that the density profiles contains mainly the
excess fermions outside the superfluid cloud at all $P$'s. This is
consistent with the homogeneous normal phase boundary extends near
resonance to large population difference \cite{pao06}. Near
resonance, the superfluid core survives at $P=0.9$ in our
calculations whereas experimentally \cite{ketterle06} it vanishes
already at $P \approx 0.7$.

On the BEC side [Fig. \ref{fig1}(g)-(i)], all of the minority are
paired up and the excess fermions can penetrate into the superfluid
core. The system contains a superfluid core for any $P$ (if
sufficiently deep in the BEC regime, see Fig. \ref{fig4} below). The
system then contains three different phases: the purely superfluid,
the mixture phase, and the normal fermions from the trap center to
the edge of the trap. The mixture phase extends toward the trap
center as the polarization increases. In Fig. \ref{fig1}(i) ($P =
0.9$), the system is highly polarized and the excess fermions extend
deeply into the trap center. In \ref{BEClimit}, we give an analytic
discussion of the density profiles in the BEC limit.

\begin{figure}
\includegraphics{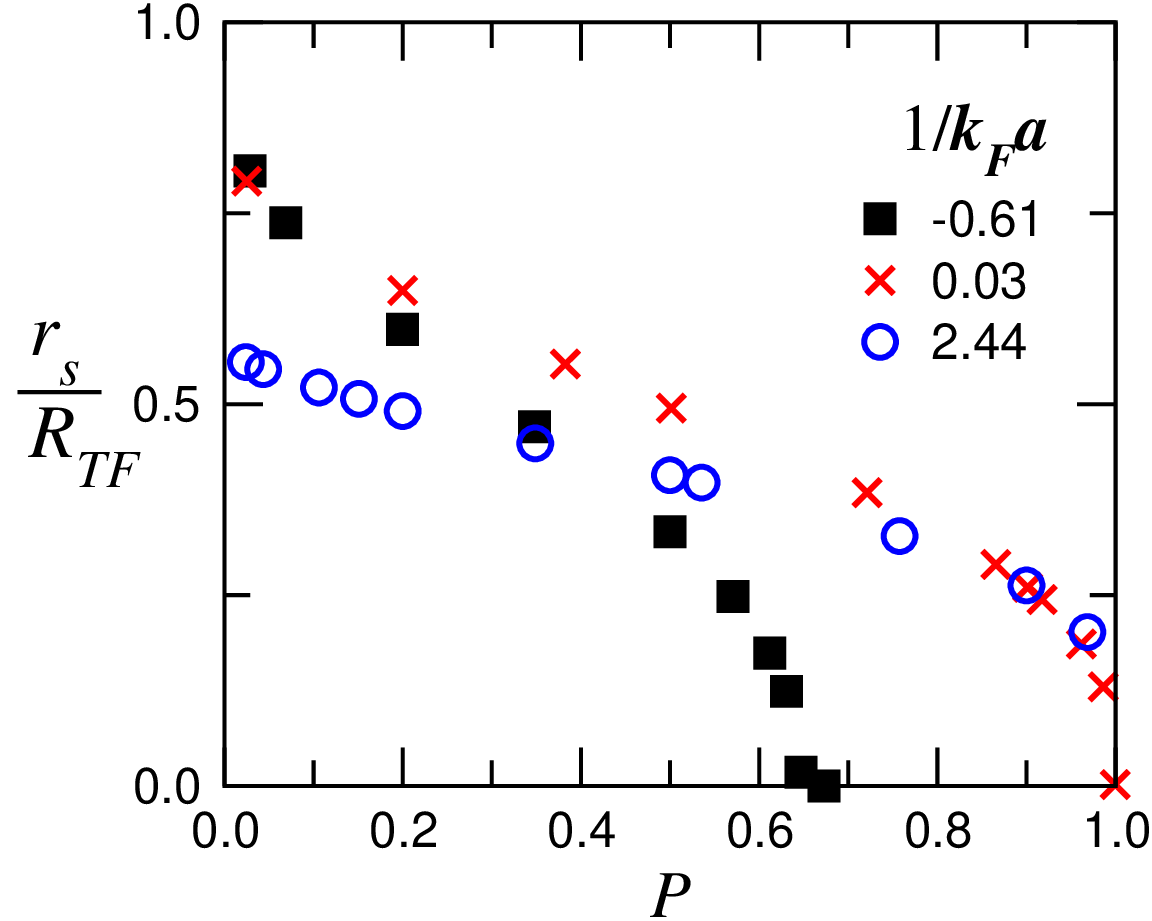} \vspace*{3.6in} \caption{(color online). Radii of the
superfluid core $r_s$ versus polarization $P$ for normalized
coupling strengths in the legend.} \label{fig2} \end{figure}

The radius $r_s$ of the superfluid core is one of the most
interesting quantities in current studies of the imbalanced fermion
system \cite{hulet06,deSilva,Haque,chevy06,kinnunen05}. From Fig.
\ref{fig1}, the density profiles of the population difference
$n_d(r)$ have maxima at the phase boundaries between the superfluid
and the normal fermions. $r_s$ is thus also the peak position of
$n_d(r)$. In Fig. \ref{fig2}, we plot $r_s$ as function of $P$ for
three different coupling strengths. $r_s$ behaves quite differently
above and below the resonance.
 On the BCS side
the superfluid is eliminated when the polarization reaches around
0.65 for $1 /(k_F a) = -0.61$ and the system becomes completely
normal with mismatch Fermi surfaces beyond this critical
polarization.
For large coupling strengths, $r_s$ is finite except when $P$ is
exactly $1$, since
 the
superfluid is stable for any finite ($\ne 1$) polarization in the
homogeneous case \cite{pao06}. Except for small ($P \le 0.1$) or
large ($P \ge 0.9$) polarizations, the sizes of the superfluid
clouds have maxima near the Feshbach resonance at fixed
polarization.

\begin{figure}
\includegraphics{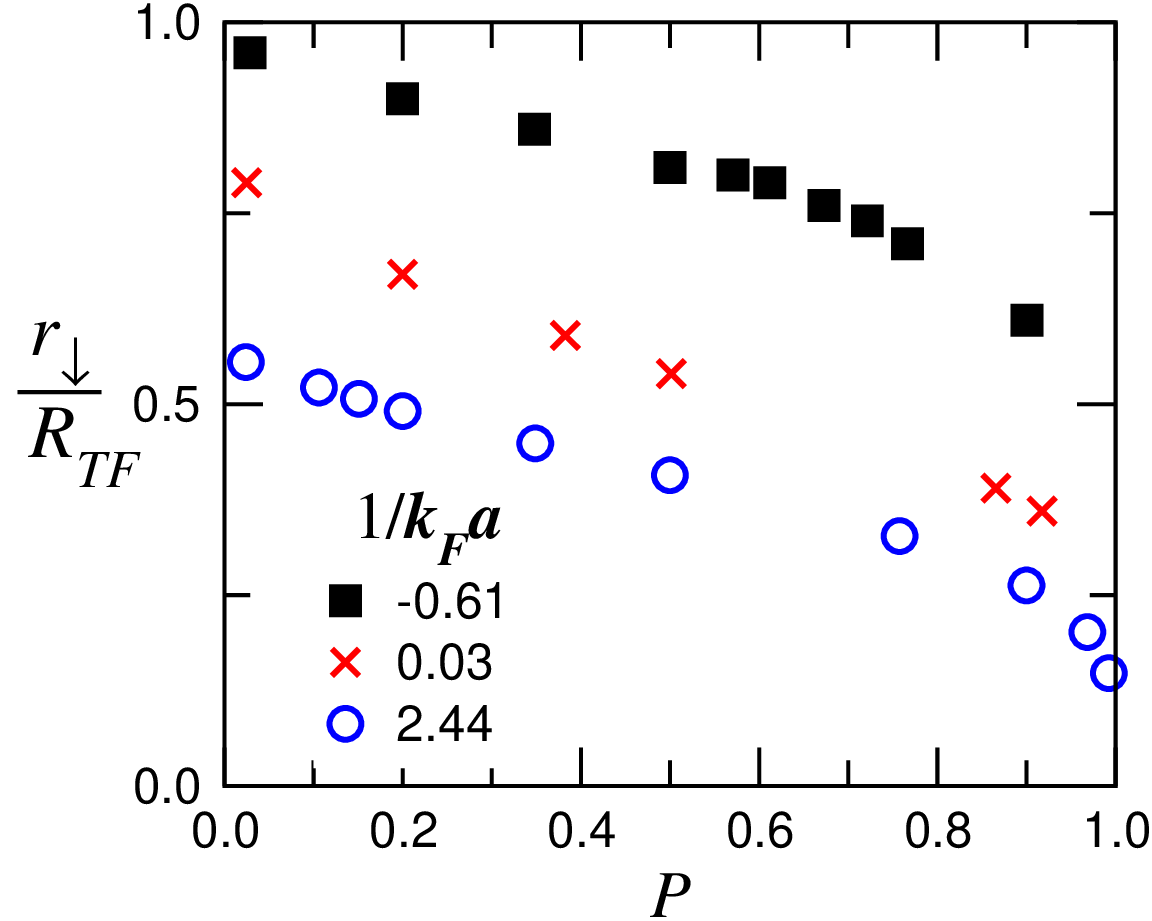} \vspace*{3.6in} \caption{(color online). Radii of the
minority species $r_{\downarrow}$ versus polarization $P$ for
normalized coupling strengths in the legend.} \label{fig3}
\end{figure}

We also plot the radii ($r_\downarrow$) of the minority (spin down
fermions) cloud in Fig. \ref{fig3}.
 Below the
Feshbach resonance, this radius is the same as the radius of the
superfluid core because all the minority of fermions are paired up.
However, this won't be true above the Feshbach resonance where part
of minority are not paired up in this regime. Unlike $r_s$,
$r_\downarrow$ decreases monotonically as the coupling strength
increases. These two radii become identical when the minority are
paired up completely when the system reaches the BEC regime.

Due to the experimental constraints, one can not measure the radial
density profiles directly. Instead the axial density profiles are
reported in  \cite{hulet06}. In Fig. \ref{fig4}, we plot the
normalized axial density profiles $n_a(z)$ $[ \equiv \int dx dy n_d
(\vec r) ]$ of the population difference for different coupling
strengths at three fixed polarization $P$. For the cases with phase
separations [Figs. \ref{fig1}(a), (b), and (d)-(f)] , the
corresponding $n_a(z)$ are constants for $z \le r_s$ and have a kink
at the phase boundary (for $z = r_s$). This feature results from the
population difference $n_d(r)$ being zero inside the superfluid
cloud (see \ref{adensity} in detail) such that $n_a (z)$ remains the
same value as at the phase boundary $z= r_s$.
 For a system with a mixed
phase region [Figs. \ref{fig1}(g)-(i)], $n_a (z)$ increases smoothly
toward the trap center even as $z \le r_s$. It reaches a constant
value within the cloud containing superfluid only [ $e.g.$, the
cases with solid lines in Figs. \ref{fig4}(a) and (b)]. At large
polarization [solid line in Fig. \ref{fig4}(c)], the excess fermions
mix with superfluid entirely for $1 / k_F a\, =\, 2.44$ such that
$n_a (z)$ increases monotonically toward the trap center. The
completely different features for the axial density profiles of the
population difference inside the superfluid cloud can help us to
clarify whether the system is phase separation or mixture.

\section{\label{conclusion}Conclusion}

We have investigated the radial density profiles of the
two-component fermion system with unequal spin-populations under
Feshbach resonance. The system shows a superfluid cloud in the trap
center surrounding by normal fermions. In the weak-coupling BCS
side, the superfluid is destroyed completely at the polarization $P
\, \gtrsim\, 0.65$ for $1 /(k_F a) = -0.61$. Near the Feshbach
resonance, almost all the minority are paired up and the system is
phase separated into superfluid and the normal fermions. In the
strong-coupling BEC side, the excess fermions can mix with the
superfluid and the system contains three different phases, the
purely superfluid cloud, the mixture phase, and the normal fermions
from the trap center to the edge of the trap.

In particular, we emphasize the difference in the axial density
difference profiles between the phase separation and Bose-Fermi
mixture regimes. The former shows a constant for $z < r_s$ and has a
kink at the phase boundary but the later are smoothly increasing
toward the trap center. However, Ref. \cite{hulet06} reported
positive slopes of the axial density different profiles inside the
superfluid cloud that indicates the $n_d (\vec r)$ need to be
negative somewhere within the superfluid cloud. We do not obtain
this phenomenon within current local density approach.

\ack This research was supported by the National Science Council of
Taiwan under grant numbers NSC94-2112-M-194-001 (CHP) and
NSC94-2112-M-001-002 (SKY), with additional support from National
Center for Theoretical Sciences, Hsinchu, Taiwan.

\appendix
\section{\label{freeenergy}Free energy}

To determine the minimum free energy state when there are multiple
solutions, we need an expression for the free energy. We obtain this
as follows.  First consider a system with fixed volume
$V$ and particle numbers $N_{\sigma}$. 
With the Hamiltonian in Eq. (\ref{eqh}), it is straight-forward to
show that \cite{AGD} the energy $E( N_{\sigma}, g) $ of this system
obeys
\begin{equation} \frac{\partial E}{\partial g} =
   \frac{V}{g^2} |\Delta|^2
\label{Eg} \ . \end{equation}
 We need to eliminate $g$ in favor of the scattering length $a$,
the physical parameter of the system.  These two variables are
related by the expression
\begin{equation}
\frac{m}{4 \pi \hbar^2 a}\ =\ \frac{1}{g}
   + \frac{1}{V} \sum_{\vec k} \frac{1}{2 \epsilon_k}\ .
\label{ag}
\end{equation}
We thus get
\begin{equation}
\frac{m}{4 \pi \hbar^2} d \left( \frac{1}{a} \right)  =
 d \left( \frac{1}{g} \right) \ ,
\label{dadg} \end{equation} and therefore \begin{equation}
\frac{\partial E}{\partial (1/a)} \vert_{N_{\sigma}} =
  - V \frac{m}{4 \pi \hbar^2} |\Delta(a) |^2 \ .
\label{Ea} \end{equation} Writing this derivative to be ${\cal E}'$,
we then get the thermodynamic relation \begin{equation} dE =
\sum_{\sigma} \mu_{\sigma} d N_{\sigma} + {\cal E}'
   d (\frac{1}{a}) \ .
\label{dE} \end{equation} The free energy $\Omega(\mu_{\sigma}, a)
\equiv
  E - \mu_{\uparrow} N_{\uparrow} - \mu_{\downarrow} N_{\downarrow} $
then obeys \begin{equation} d \Omega = - \sum_{\sigma} N_{\sigma} d
\mu_{\sigma} +
   {\cal E}'   d (\frac{1}{a}) \ .
\label{dOmega} \end{equation} We thus conclude that, for volume $V$
and chemical potentials $\mu_{\sigma}$, \begin{equation}
\frac{\partial \Omega}{\partial (1/a)} \vert_{\mu_{\sigma}} =
  - V  \frac{m}{4 \pi \hbar^2} |\Delta|^2 \ .
\label{dOda} \end{equation} Though this expression is already
sufficient to determine and thus compare the free energies of the
different solutions for given scattering length $a$, we can convert
it to an even more convenient form. To do this, let us write $x
\equiv |\Delta|^2$ and $y \equiv 1/a$. We get, up to an overall
multiplying factor
\begin{equation}
\frac{\partial \Omega}{\partial y} =  -  x \ .\label{dOdx}
\end{equation} Thus, when the solutions are plotted in the form of
Fig. \ref{fig5}, the free energy $\Omega$ of a state can be related
to the free energy $\Omega_0$ of another state along the same curve
by, again up to an overall multiplicative constant,
\begin{equation} \Omega = \Omega_0 - \int x dy \ .\label{Oxy} \end{equation}
Since $\int x dy$ is the area between the curve and the y-axis, the
states corresponding to the minimum free energy at a given
scattering length $a$ (hence $y$) can be found via the same
procedure as the usual Maxwell construction.

\section{\label{BEClimit}BEC limit}

 Here we try to understand the behaviour of the density profile in the BEC
($ 1/k_F a \gg 1$) limit, for the special case of small number of
excess fermions [$e.g.$ Fig. \ref{fig1}(g)]. For a bulk system, it
is straight forward to perform a low density expansion of the
mean-field equations and obtain an expansion of the chemical
potentials $\mu_f \equiv \mu_{\uparrow}$ and $\mu_b \equiv
\mu_{\uparrow} + \mu_{\downarrow}$ as a series in the densities $n_f
= n_d$ and $n_b \equiv n_{\downarrow}$ (see \cite{Yip02} for the
details of this calculation). In the BEC limit, $n_d$ can be
interpreted as the density of the excess unpaired fermions and $n_b$
as the density of the bosons which represent the bound fermion
pairs. $\mu_f$ and $\mu_b$ can be interpreted as the corresponding
chemical potentials.  Explicitly we have an expansion of the form
\begin{eqnarray}
\mu_f &=& A n_d^{2/3} + g_{bf} n_b \ ,
\label{muf} \\
\mu_b &=& - \epsilon_b + g_{bb} n_b + g_{bf} n_f \ , \label{mub}
\end{eqnarray}
where we have dropped the terms higher order in the densities. These
terms are smaller in the limit $n_f a^3 \ll 1$ and $n_b a^3 \ll 1$.
We obtain \cite{Yip02} $A = (6 \pi^2)^{2/3} / 2 m$, $\epsilon_b =
\hbar^2 / m a^2$, $g_{bb} = 4 \pi \hbar^2 a / m$, $g_{bf} = 8 \pi
\hbar^2 a / m$. The value of $A$ is as expected for a free Fermi gas
and $\epsilon_b$ is the binding energy of the fermion pair. The
values of $g_{bb}$ and $g_{bf}$ here, obtained from the mean-field
equations, differ from the correct values known. Exact three-body
calculation \cite{Skorniakov} gives $g_{bf}^{\rm exact} = 3.6 \pi
\hbar^2 a / m$, while a recent four-body calculation \cite{Petrov04}
gives $g_{bb}^{\rm exact} = 1.2 \pi \hbar^2 a / m$. Alternatively,
defining the scattering lengths in the usual manner, our low density
expansion yields the effective values $a_{bf} = 8 a /3$ and $a_{bb}
= 2 a$, whereas the exact values should be $a_{bf}^{\rm exact} = 1.2
a$ and $a_{bb}^{\rm exact} = 0.6a $. The difference between the
mean-field and exact results is due of course to the approximate
nature of the mean-field theory. We note however, here that (a)
$a_{bf}$ and $a_{bb}$ are both positive and of order $a$, hence in
the BEC limit we have necessarily $n_d a^3 \ll 1$, thus the fermions
and the bosons can mix \cite{viverit00}; (b) $g_{bf}/ g_{bb} > 1/2$
for both mean-field and exact values.  We shall explain the
importance of this second relation below.

In the trap, Eqs. (\ref{muf}) and (\ref{mub}) becomes
\begin{eqnarray}
\mu_f &=& A n_d (\vec r) ^{2/3} + g_{bf} n_b (\vec r) + V( \vec r) \
,
\label{muft} \\
\mu_b &=& - \epsilon_b + g_{bb} n_b (\vec r) +
   g_{bf} n_f (\vec r)+ 2 V (\vec r) \ .
\label{mubt}
\end{eqnarray}
Here $V(\vec r)$ is the trap potential.  We shall only consider the
case where the potential increases from the center of the trap. To
appreciate the implications of the relation (b), consider the limit
of a small number of fermions.  The density profile of the bosons
can then be determined by first ignoring the fermion term in Eq.
(\ref{muft}), and we obtain the boson density profile
\begin{equation} n_b (\vec r) = (\mu_b + \epsilon_b - 2 V (\vec
r))/g_{bb} \ ,\label{nb0} \end{equation} if the R.H.S. is larger
than zero, and $n_b = 0$ otherwise. For ease of reference, we shall
call these two regions ``inside" and ``outside" below.

Eq. (\ref{muft}) can be rewritten in the form
\begin{equation} \mu_f
= A n_d (\vec r)^{2/3} + V_{\rm eff} (\vec r) \ ,\label{eff}
\end{equation}
where $V_{\rm eff} (\vec r) = V (\vec r) + g_{bf} n_b
(\vec r)$ is the effective potential for the fermions. We obtain,
for the inside region, $V_{\rm eff} (\vec r) = V (\vec r)
  + (g_{bf} /g_{bb}) (\mu_b + \epsilon_b - 2 V (\vec r))$
whereas for the outside $V_{\rm eff} (\vec r) = V (\vec r)$. In the
inside region, the position dependence is therefore $  - (2
g_{bf}/g_{bb} -1 ) V(\vec r)$. From this, it is clear that if
$g_{bf} > g_{bb} /2$, then the effective potential actually is
larger near $\vec r = 0$ than near the edge of the boson cloud.  It
decreases from the center of the trap till it reaches the edge of
the boson cloud, then it increases again due to the spatial
dependence of $V (\vec r)$.  Therefore, we conclude that for
${g_{bf}}/{g_{bb}} > 1/2$, the excess fermions lie near the edge of
the boson cloud, at least for a small number of Fermions
\cite{Carr04,pieri05}. The excess fermions lie near the edge of the
cloud [Fig. \ref{fig1}(g)]. 
Thus a peak in $n_d(\vec r)$ does {\it not} indicate phase
separation.

\section{\label{adensity}Axial Density}

We here discuss some useful expressions governing the axial density
within the local density approximation.  Our results here are, to
some extent, a further development of those obtained in
\cite{deSilva}.

In local density approximation, any quantity, say the density
difference $n_d (\vec r)$, is a function entirely of the local
chemical potential $\mu (\vec r)$ (note that $h$ is a constant).
Hence, we can write
\begin{equation}
n_d (\vec r) = g (\mu(\vec r))
\equiv g (\mu(\vec r), h) \label{g}
\end{equation}
for some function $g$. We here note that any density must be
identically zero for sufficiently large and negative chemical
potential, and thus $g(\zeta)$ vanishes exactly for sufficiently
large and negative argument $\zeta$. It will be convenient to define
another function $G(\zeta)$ via
\begin{equation}
G(\zeta) \equiv \int_{-\infty}^\zeta d\zeta' g(\zeta') \ .\label{G}
\end{equation}

Consider now for example the radially integrated density (referred
afterwards as axial density) difference:
\begin{equation} n_{a}
(z) \equiv \int dx dy n_d (\vec r) \ .\label{Na} \end{equation} The
integral can be written as, for a trap that is cylindrically
symmetric with respect to $z$, $\pi \int_0^{\infty} d \rho^2 g
(\mu_0 - \frac{1}{2} \alpha_z z^2 - \frac{1}{2} \alpha_{\parallel}
\rho^2 ) $ where $\rho^2 \equiv x^2 + y^2$. This integral can be
expressed in terms of $G$ and thus
\begin{equation}
n_{a}(z) = \frac{2 \pi}{\alpha_{\parallel}} G (\mu_0 - \frac{1}{2}
\alpha_z z^2) \ .\label{NaG} \end{equation}

This relation can be used to obtain directly the axial density in
terms of the function $g$, without first calculating the density
profile in trap and then performing the integration. Moreover, it
can also be used to deduce the (unintegrated) density profile once
the axial density is given, for example from experiment. To see
this, we differentiate Eq. (\ref{NaG}) with respect to $z$, and
notice that $G'(\zeta) = g(\zeta)$ evaluated at $\zeta = \mu_0 -
\frac{1}{2} \alpha_z z^2 $ is directly related to the corresponding
density at the coordinate $(0,0,z)$. Thus we have
\begin{equation}
n_d(0,0,z) = - \frac{\alpha_{\parallel}}{2 \pi \alpha_z} \frac{1}{z}
\frac{\partial n_{a}(z)}{\partial z}\ . \label{nz}
\end{equation}
Therefore the actual density for a point on the $z$ axis can be
obtained from the axial density.  Under the local density
approximation, the density at any given point in the trap is a
function of the local chemical potential only. Hence we can obtain
the actual density at any given point in space provided the axial
density is given.

The relation Eq. (\ref{nz}) also shows that, since $n_d$ is
non-negative, the axial density profile is strictly decreasing
(increasing) with $z$ for $z\, >\, (<)\, 0$, a result pointed out in
reference \cite{deSilva}. In a region where the density vanishes,
then the axial density should be a constant, a result recognized in
references \cite{deSilva} and \cite{Haque}.

\newpage
\noappendix

\newpage

\newpage
\begin{figure}
\includegraphics{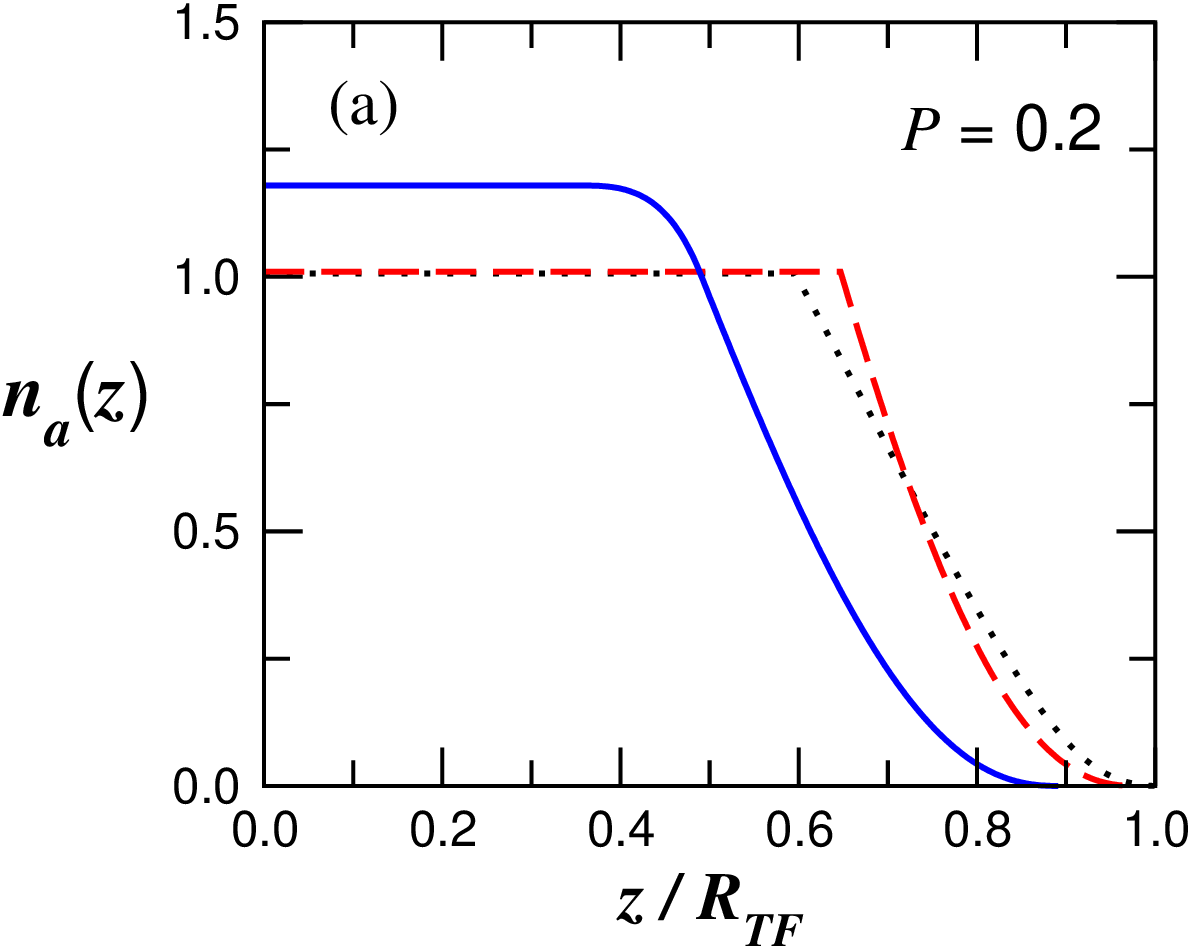} \vspace*{2.8in} \includegraphics{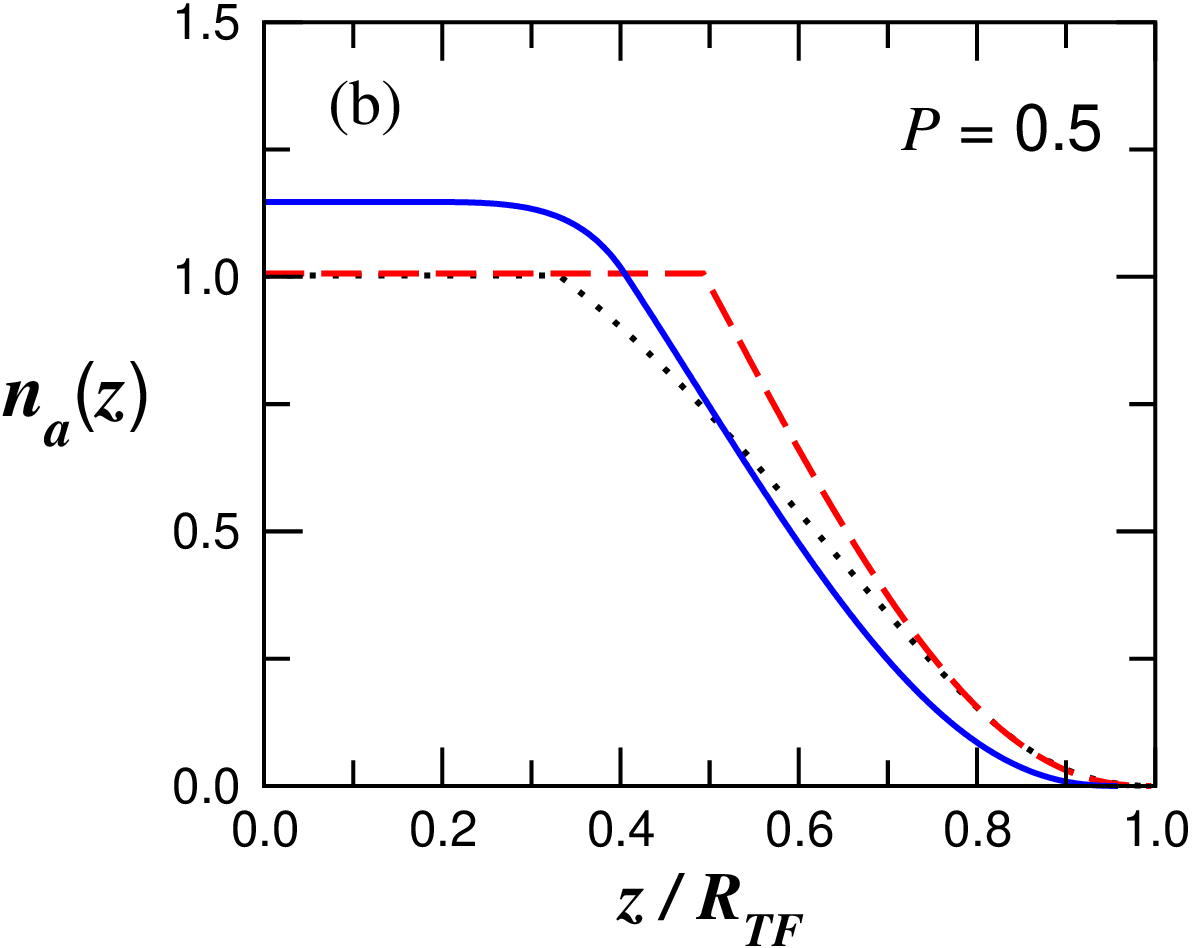}\vspace*{2.8in}
\includegraphics{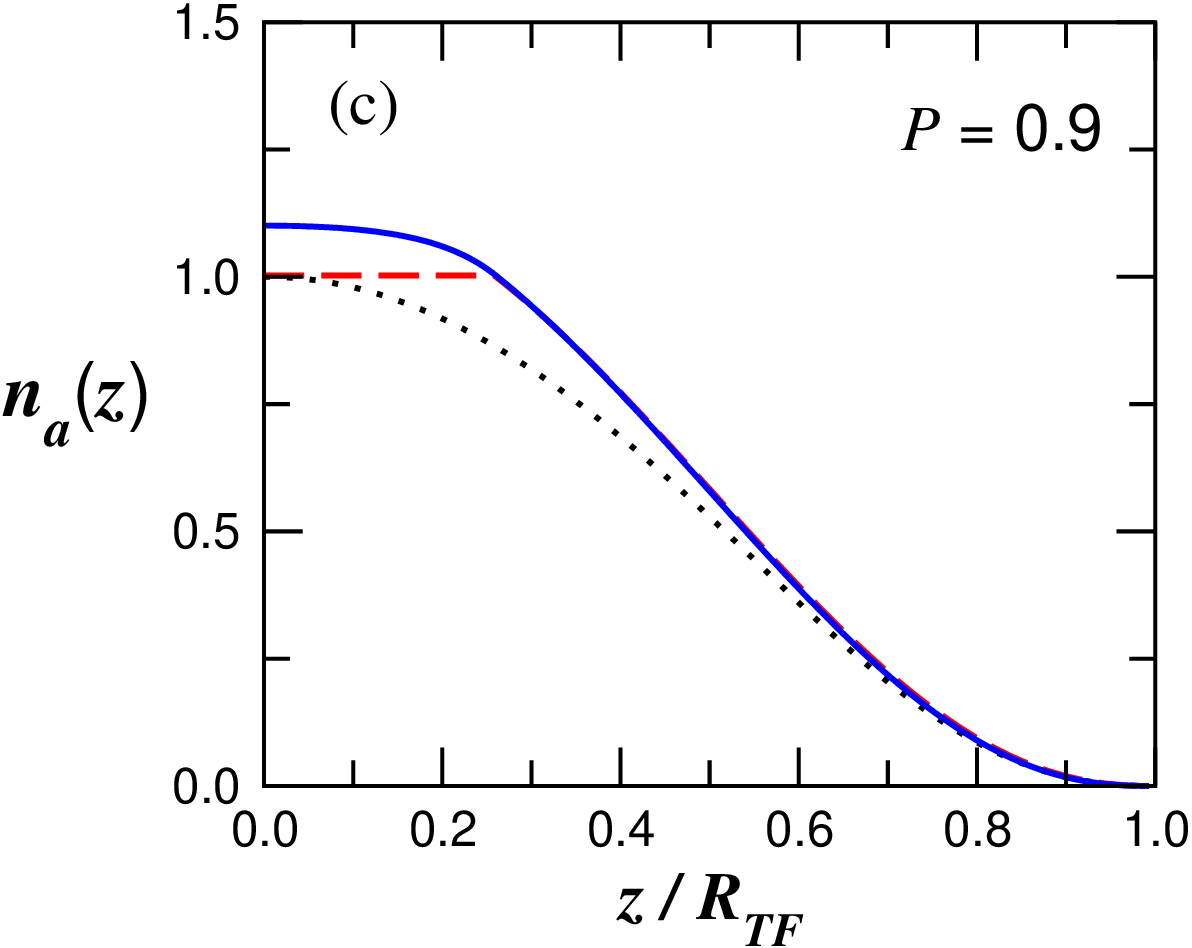} \vspace*{2.9in} \caption{(color online). Axial
population density difference $n_a(z)$ versus $z/R_{TF}$. $R_{TF}$
is defined in the caption of Fig 1. $n_a (z)$ is normalized to $1$
at $z = r_s$.} \label{fig4}
\end{figure}

\newpage

\begin{figure}
\includegraphics{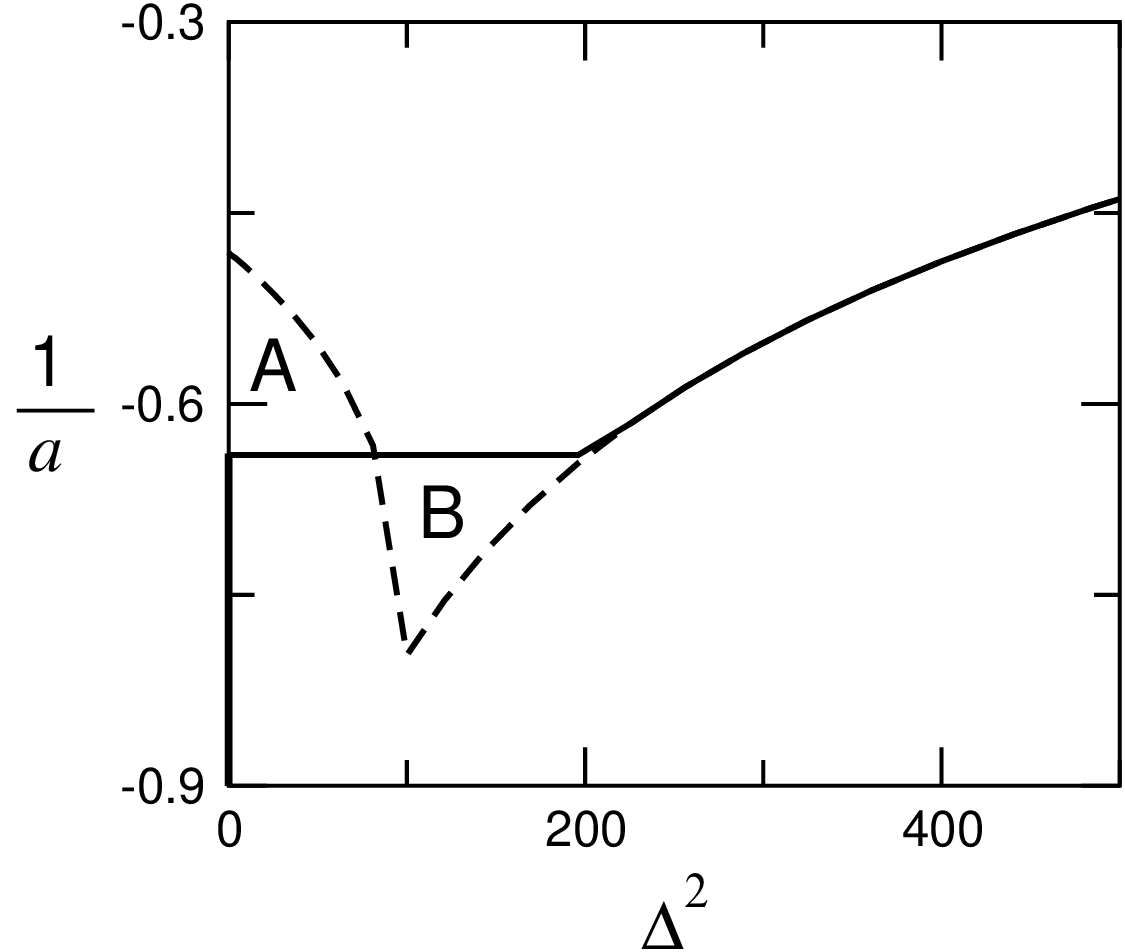} \vspace*{3.2in} \caption{Solution to the gap equation
plotted as $1/a$ vs $|\Delta|^2$ (arbitrary units) for the case with
multiple solutions (dashed lines).  The minimum energy solution is
shown as the thick black line.  The areas labeled by A and B are
equal.} \label{fig5}
\end{figure}

\end{document}